\documentclass[conference]{IEEEtran}
\IEEEoverridecommandlockouts
\PassOptionsToPackage{hyphens}{url}\usepackage{hyperref}
\usepackage{cite}
\usepackage{amsmath,amssymb,amsfonts}
\usepackage{algorithm,algpseudocode}
\usepackage{graphicx}
\usepackage{mdframed}
\usepackage{multirow}
\usepackage{textcomp}
\usepackage{xcolor}
\usepackage{url}
\usepackage{listings}
\def\BibTeX{{\rm B\kern-.05em{\sc i\kern-.025em b}\kern-.08em
    T\kern-.1667em\lower.7ex\hbox{E}\kern-.125emX}}

\begin{document}

\newtheorem{hypothesis}{Hypothesis}

\definecolor{dkgreen}{RGB}{0, 128, 0}

\title{Impact of Comments on LLM Comprehension of Legacy Code}

\author{\IEEEauthorblockN{Rock Sabetto, Emily Escamilla, Devesh Agarwal, Sujay Kandwal, \\Dr. Justin F. Brunelle, Dr. Scott Rosen, Dr. Nitin Naik, Dr. Samruddhi Thaker, \\Dr.\ Eric O.\ Scott, Jacob Zimmer, Amit Madan, Arun Sridharan, Doug Wendt,\\ Michael Doyle, Christopher Glasz, Jasper Phillips, William Macke,\\ Colin Diggs, Michael Bartholf, Zachary Robin, and Paul Ursino \textsuperscript{\textsection}}
\IEEEauthorblockA{\textit{The MITRE Corporation} \\
McLean, VA\\
rsabetto@mitre.org}
}

\maketitle

\begingroup\renewcommand\thefootnote{\textsection}
\footnotetext{Due to the size of the team, we list everyone here but see the first author's contact information below as the corresponding author}
\endgroup

\begin{abstract}

Large language models (LLMs) have been increasingly integrated into software engineering and maintenance tasks due to their high performance with software engineering tasks and robust understanding of modern programming languages. However, the ability of LLMs to comprehend code written with legacy languages remains a research gap challenged by real-world legacy systems lacking or containing inaccurate documentation that may impact LLM comprehension. To assess LLM comprehension of legacy languages, there is a need for objective LLM evaluation. In order to objectively measure LLM comprehension of legacy languages, we need an efficient, quantitative evaluation method. We leverage multiple-choice question answering (MCQA), an emerging LLM evaluation methodology, to evaluate LLM comprehension of legacy code and the impact of comment prevalence and inaccurate comments. In this work, we present preliminary findings on the impact of documentation on LLM comprehension of legacy code and outline strategic objectives for future work.

\end{abstract}

\begin{IEEEkeywords}
artificial intelligence, large language models, legacy software, software modernization, multiple-choice question answering
\end{IEEEkeywords}

\section{Introduction}

With the widespread adoption of large language models (LLMs), these models are increasingly being utilized for software development and maintenance tasks, including generating code \cite{chen2021evaluating,fan2023large}, creating code comments \cite{dvivedi2024comparative,khan2022automatic}, and debugging code\cite{majdoub2024debugging,tian2024debugbench}. Effectively employing LLMs for these tasks is contingent on the ability of LLMs to understand code. The extensive availability of open-source data for modern programming languages like Python, Java, and C++ has likely contributed to the high performance of LLMs in code-related tasks \cite{leinonen2023comparing,khan2022automatic,nam2024using}, reinforcing the assumption that LLMs can effectively comprehend modern code languages. However, legacy languages like assembly, COBOL, and Fortran are inherently less prevalent, which may result in limited training data for LLMs. As developers deploy LLMs for tasks related to software maintenance and modernization for legacy languages, it is crucial they ensure that LLMs possess a sufficient understanding of the relevant subject matter to deliver high-quality outputs.

In real-world software systems, comments serve as a crucial form of documentation for understanding code and code updates. However, the extent and quality of documentation can vary significantly, and existing documentation may deteriorate in accuracy over the system's lifecycle. For instance, code modifications made to address reported bugs may not be accompanied by corresponding updates to the comments. Because input data impacts the quality of outputs when utilizing LLMs \cite{fan2023large}, lack of comments or inaccuracy of comments may negatively impact LLM comprehension of the provided legacy code, limiting the application of LLMs in accelerating code modernization. This study examines the influence of documentation, particularly code comments, on the comprehension of legacy programming languages by LLMs. This investigation is guided by two research questions:

\begin{itemize}
    \item \textbf{RQ1:} Does the prevalence of accurate comments improve LLM comprehension of legacy code?
    \item \textbf{RQ2:} Do inaccurate code comments degrade LLM comprehension of legacy code?
\end{itemize}

To answer these research questions, we utilize multiple-choice question answering (MCQA), a novel method of LLM evaluation, to measure LLM comprehension of legacy code languages. 

\section{Related Work}

The application of LLMs to software engineering tasks for modern languages has been extensively researched in recent years \cite{fan2023large}. Research has shown that LLMs are able to create high quality documentation that matches or exceeds the quality of human-generated documentation for Python \cite{dvivedi2024comparative} and a wide range of modern languages \cite{khan2022automatic}. Beyond summarizing code, LLMs have been successful in generating code \cite{chen2021evaluating} and debugging code \cite{majdoub2024debugging,tian2024debugbench} for modern languages including C++, Python, and Java. These findings indicate extensive LLM comprehension and knowledge of modern languages. However, the application of LLMs to software engineering and maintenance tasks for legacy languages is an existing research gap with their performance being uncertain although showing initial promise in generating comments \cite{pietrini2024bridging,diggs2024leveraging}.

The assessment of free-form LLM outputs as the traditional approach to LLM evaluation is challenging and evaluations based on text matching are insufficient for assessing LLM performance \cite{wang2024llms}. Evaluating LLMs with MCQA is an emerging methodology across many domains \cite{li2024can,wang2024llms} including software engineering \cite{zhang2024multiple}. MCQA is non-subjective and easily automated \cite{wang2024llms}, allowing for simple and quick assessments of model performance \cite{li2024can}. With MCQA, LLMs are prompted to answer multiple choice questions and answer accuracy provides a quantitative measure of model performance. Further demonstrating the momentum around MCQA, the popular benchmarks MATH, HumanEval, GSM8K, and MBPP have been recreated in the form of multiple-choice benchmarks \cite{zhang2024multiple}. Due to the identified research gap of LLM comprehension of legacy languages and the promise of MCQA as an emerging evaluation methodology, we explore the use of MCQA to measure LLM comprehension and the impacts of comment frequency and accuracy on LLM comprehension. 

\section{Methods}


To measure the impact of comment prevalence and inaccurate comments on LLM comprehension of legacy code, we developed a methodology to utilize MCQA. We identified a representative code sample for our experimentation, crafted multiple-choice question (MCQ) quizzes with the help of subject matter experts (SMEs), and measured comprehension LLM based on MCQA accuracy as outlined in the following three subsections. 

\subsection{Experimental Design}

For all of our experiments, we used L8WAIT\footnote{\url{https://github.com/walmartlabs/zFAM/blob/master/Source/L8WAIT.asm}}, a mainframe assembler language module from an open-source codebase published by Walmart \cite{Walmart2024} that applies conditional logic to establish a response wait time. The Walmart codebase is written in IBM Z/OS, a version of assembler used in multiple operational legacy government systems that has proven challenging to modernize. Therefore, we selected the L8WAIT module as the basis of our experiment as a representative sample of real-world legacy code.

To assess the impact of comment prevalence for \textbf{RQ1}, we created four variations of the L8WAIT module with various levels of comments: code with all comments, code with only function-level comments\footnote{In assembler, line-wise comments are called comments while function-level comments are called remarks. For generalizability, we will refer to remarks as function-level comments and comments as line-wise comments}, code with only line-wise comments, and code with no line-wise or function-level comments. 

To assess the impact of inaccurate comments, we created additional two variations of the L8WAIT module. For the first variation, we replaced 20\% of the comments with inaccurate comments. We changed comments that described L8WAIT time delay functionality to comments describing how to calculate a tip for restaurant waitstaff. With this variation we sought to measure the impact of minimal comment inaccuracies on the LLMs' understanding of the code.  For the second variation, we wrote a logical description of a National Basketball Association (NBA) simulation and used it in place of all L8WAIT comments and remarks to measure the impact of widespread comment inaccuracies.

\subsection{Multiple-Choice Quiz Creation}

We leveraged MCQA as a proxy for measuring LLM comprehension of legacy code and, specifically, LLM understanding of the L8WAIT module. We determined degraded quiz accuracy to be a reflection of degraded LLM comprehension of legacy code and, inversely, improved quiz accuracy to be a reflection of improved comprehension. In consultation with two assembler language SMEs, we developed and vetted a total of four quizzes to establish and measure LLM understanding of the L8WAIT module. A summary of the research questions and associated L8WAIT variations and quizzes is shown in Table \ref{tab:experiments}.

\begin{figure}
    \centering
    \includegraphics[width=\columnwidth]{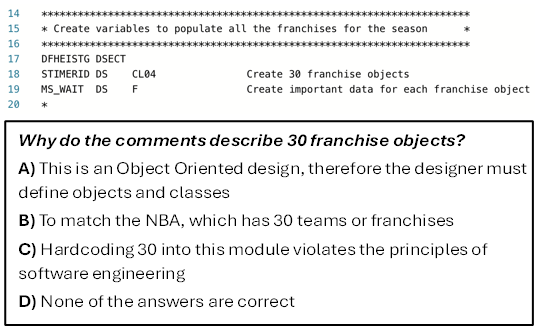}
    \caption{An example multiple-choice question and the code it is referring to. The question is one of 10 designed to gauge LLMs' ability to detect that the code does not correlate with the comments. The comment in line 18 refers to creating franchise objects, but the code establishes the timer ID. Therefore, the correct answer is \textbf{D}.}
    \label{fig:example_experiment_3_question}
\end{figure}

To measure the impact of comment prevalence, we created a basic and an advanced quiz. The basic quiz consisted of 20 basic-level multiple-choice questions focused on overall program function and command meaning. The advanced quiz consisted of 37 more challenging multiple-choice questions regarding program function, command meaning, and data processing. To measure the impact of inaccurate comments we adapted the advanced quiz to more specifically gauge the influence of inaccurate comments on overall program understanding. For the variation with 20\% inaccurate comments, we replaced 25\% of the questions from the advanced quiz with questions regarding comment inaccuracy, resulting in a 37-question quiz. For the variation with completely inaccurate comments, we created a 20-question quiz: 10 questions selected from the advanced quiz covering program function and command meaning and 10 questions created to assess LLM ability to detect inaccurate context embedded within comments. An example of a question targeting comment accuracy is shown in Figure \ref{fig:example_experiment_3_question}. 

\begin{table}[]
\caption{Layout of the research questions and associated L8WAIT variations and quizzes. All combinations of L8WAIT variations and quizzes were applied to all four LLMs used in this experiment (Claude-3, Llama 3, Mixtral, and GPT-4) as described in Section \ref{sec:measuring}.}
\label{tab:experiments}
\resizebox{\columnwidth}{!}{%
\begin{tabular}{lll}
\textbf{RQ} & \textbf{L8WAIT Variations} & \textbf{Quizzes} \\ \hline
\multirow{2}{*}{RQ1} &
  \begin{tabular}[c]{@{}l@{}}No comments, Function-level comments, \\ Line-wise comments, All comments\end{tabular} &
  Basic (20 questions) \\ \cline{2-3} &
  \begin{tabular}[c]{@{}l@{}}No code, No comments, \\ Function-level comments, \\ Line-wise comments, All comments\end{tabular} &
  Advanced (37 questions) \\ \hline
\multirow{2}{*}{RQ2} &
  20\% Inaccurate Comments &
  \begin{tabular}[c]{@{}l@{}}Modified Advanced Quiz\\ (37 questions)\end{tabular} \\ \cline{2-3} & All Inaccurate Comments    & \begin{tabular}[c]{@{}l@{}}Modified Advanced Quiz\\ (20 questions)\end{tabular} \\ \hline
\end{tabular}
}
\end{table}

\subsection{Measuring Comprehension}
\label{sec:measuring}

To measure the impact of comment prevalence and comment inaccuracies on LLM comprehension, we provided four LLMs (Claude-3, Llama 3, Mixtral, and GPT-4) with each of the six L8WAIT variations and prompted them to respond to the associated quiz or quizzes. To answer \textbf{RQ1}, we provided each of the four variations of the L8WAIT module concerning comment prevalence to the LLM and prompted it to respond to both the basic quiz and the advanced quiz. We repeated this procedure three times for each LLM to ensure consistency and reliability in the results.

To preemptively address concerns regarding the inclusion of the L8WAIT module in the LLMs' pre-existing training data, we performed another set of quiz iterations. In these iterations, the LLMs were not provided with the L8WAIT code and were prompted to answer the advanced quiz. This process was similarly repeated three times for each LLM.

To answer \textbf{RQ2}, we provided the LLM with the L8WAIT variation with 20\% inaccurate comments and prompted it to respond to the associated 37-question quiz. We then provided the LLM with the L8WAIT variation with completely inaccurate comments and prompted it to respond to the associated 20-question quiz. This process was repeated three times for each of the four LLMs to maintain consistency across experiments. 

\section{Results and Discussion}

\begin{figure*}
    \centering
    \includegraphics[width=\textwidth]{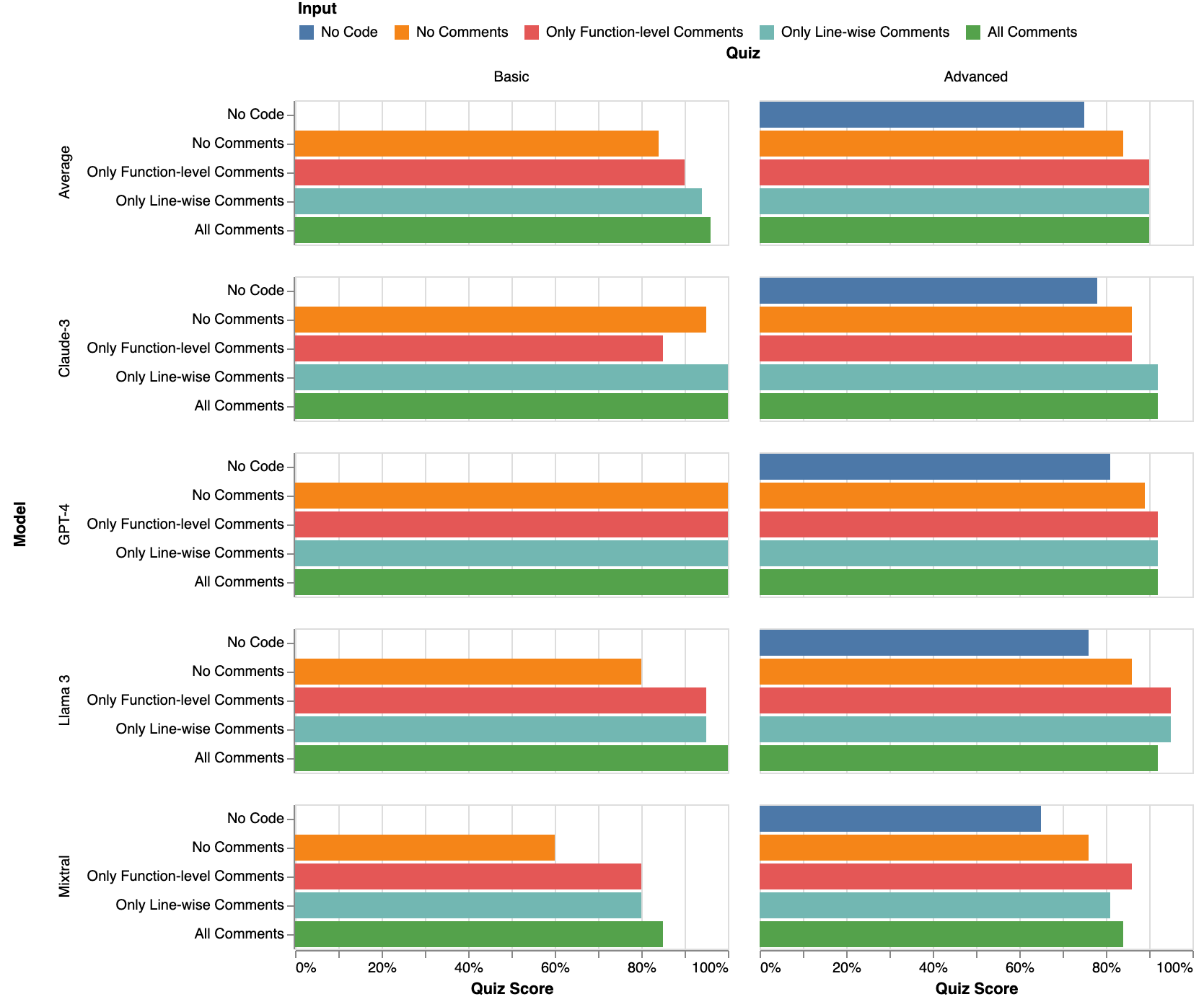}
    \caption{Scores for both the basic and advanced quizzes from all four LLMs when provided with the four L8WAIT module variations related to comment prevalence. \textit{Note:} The ``No Code'' variation was only tested with the advanced quiz as described in Section \ref{sec:measuring}}
    \label{fig:comment_prevalence}
\end{figure*}

We begin by analyzing the results related to \textbf{RQ1} to determine how the prevalence of comments affects LLM comprehension, as shown in Figure \ref{fig:comment_prevalence}. To determine the impact of comment prevalence on LLM comprehension, we averaged the quiz scores from each of the three runs to calculate an average quiz score for the given combination of input, quiz type, and LLM. In all cases, the presence of comments resulted in increased quiz scores (96\% for the basic quiz and 90\% for the advanced quiz) when compared to quiz scores for the L8WAIT module variation with no comments (84\% for both the basic and advanced quizzes). While GPT-4 had little variance in scores across variants, Llama 3 and Mixtral benefited the most from additional comments when responding to the basic quiz. Llama 3's performance improved from a score of 80\% with no comments to 100\% with all comments and Mixtral's performance improved from a score of 60\% with no comments to 85\% with all comments. Generally, increasing the number of comments provided to the LLM resulted in higher quiz scores. Additionally, we see that providing either function-level comments or line-wise comments results in higher quiz scores than when no comments are provided. 

In this study, we propose that MCQA can be used as a proxy to measure LLM comprehension. Therefore, in response to \textbf{RQ1}, \textit{we conclude that increasing comment prevalence results in increased LLM comprehension.} In most cases, the highest levels of LLM comprehension were a result of extensive comments. In application, this finding promotes the use of well-documented code in complex code-related LLM tasks.

Figure \ref{fig:comment_prevalence} also shows the results of prompting the LLM to respond to the advanced quiz without providing any code. When not provided with the code, the LLMs achieved an average score of 75\%. Therefore, it is likely that the L8WAIT module exists in the training data. However, despite the LLMs having some knowledge of the module, simply providing the code \textit{without} comments resulted in increased LLM comprehension reflected in an average score of 84\% across LLMs.

\begin{figure}
    \centering
    \includegraphics[width=0.7\columnwidth]{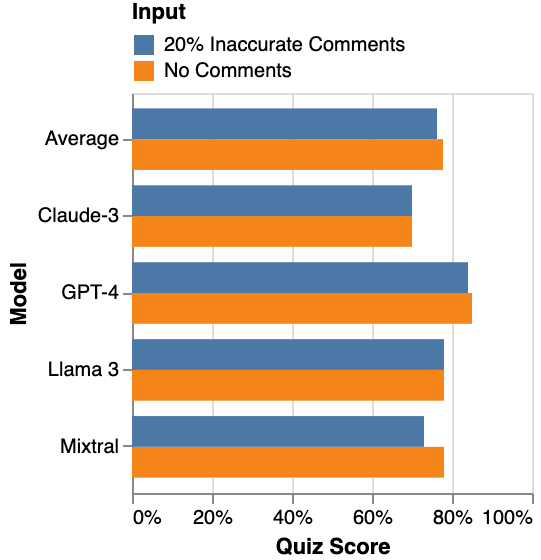}
    \caption{Quiz scores for all four LLMs when provided with the L8WAIT module variation with 20\% inaccurate comments and the variation with no comments. LLMs were prompted to answer the modified advanced quiz.}
    \label{fig:minimal}
\end{figure}

\begin{figure}
    \centering
    \includegraphics[width=\columnwidth]{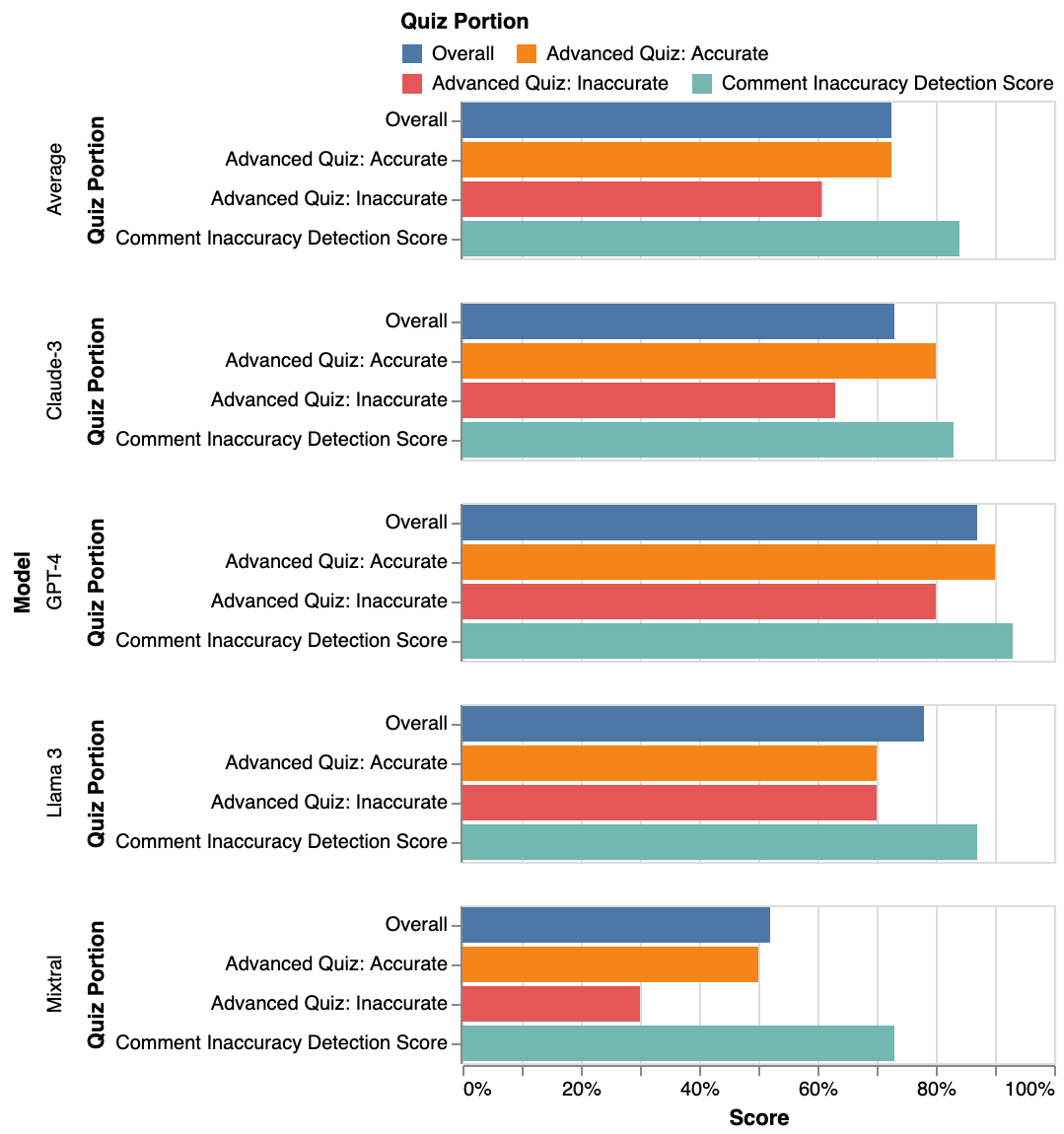}
    \caption{Performance comparison of the four language models on a quiz comprising two sections: the questions from the Advanced Quiz and Comment Inaccuracy Detection questions. The Overall score represents the cumulative performance on both sections. The Advanced Quiz score is divided into scenarios where the models were provided with accurate comments (orange) and inaccurate comments (red). The Comment Inaccuracy Detection score (green) evaluates the models' ability to identify inaccuracies in comments that do not match the code.}
    \label{fig:complete}
\end{figure}

Next, we analyzed the results related to \textbf{RQ2} to determine how inaccurate comments affect LLM comprehension, as shown in Figures \ref{fig:minimal} and \ref{fig:complete}. To determine the impact of inaccurate comments on LLM comprehension of legacy code, we averaged the quiz scores for each of the three runs to create an average score for the unique combination of LLM and L8WAIT module variation. As shown in Figure \ref{fig:minimal}, replacing 20\% of the comments in the L8WAIT module with inaccurate comments had negligible effect on quiz scores. Claude-3, GPT-4, and Llama 3 seemed to disregard the inaccurate comments and achieved identical scores on the quiz when provided first with the L8WAIT module with no comments and, second, when provided with the inaccurate context. However, as shown in Figure \ref{fig:complete}, replacing all comments with inaccurate comments noticeably impacted quiz scores with an average score of 73\% across the LLMs. The LLMs scored an average of 73\% on the 10 questions selected from the advanced quiz when given the original, accurate comments. The quiz scores degraded to an average of 61\% on the same questions when given completely inaccurate comments. Additionally, as measured by the ``Comment Inaccuracy Detection'' portion of the quiz, LLMs were only able to achieve an average score of 84\% when asked targeted multiple choice questions about the accuracy of the comments within the code. 

Given the clear degradation of LLM comprehension when provided with completely inaccurate comments, we sought to further understand the LLM’s understanding of the relationship between the L8WAIT code and the NBA simulation description. Each LLM was prompted to provide a short description of the functionality of the L8WAIT module variation containing the NBA simulation in the comments and remarks. Here, some of the LLMs conflated the relationship between the code and the comments to the point of generating false conclusions. In particular, Mixtral responded that the L8WAIT module, ``includes various subroutines to handle different aspects of the simulation, such as determining talent levels, generating a schedule, adding free agents to teams''. GPT-4 stated, ``This program is a comprehensive simulation tool for predicting the outcome of an NBA season''. In response to \textbf{RQ2}, the results of our experimentation surrounding the impact of inaccurate comments on LLM comprehension of legacy code show that, \textit{while minor inaccuracies are largely tolerated by LLMs, major inaccuracies result in 12\% degradation of LLM comprehension even when the LLM has the code in question in its training dataset.} Even when the LLM can be expected to know the answer, misplaced or incorrect comments significantly degrade the LLM's ability to comprehend the operation of legacy languages. As opposed to humans who would often come to the clear conclusion that the comments do not match the code, the LLMs believe that all provided input is true and correct and works to rectify the provided input to the point of conflating the relationship between given information.

\section{Conclusion}

In this study, we applied an MCQA experimentation framework to explore the impact of code comments on the comprehension of legacy programming languages by LLMs. Through a series of experiments using the L8WAIT module, we examined how the presence and accuracy of comments influence LLM performance in understanding legacy code. With \textbf{RQ1}, we investigated the relationship between the prevalence of code comments and LLM comprehension of legacy code. Our findings indicate that increased comment prevalence improves LLM comprehension, as evidenced by improved quiz accuracy when comments are present. This suggests that well-documented code enhances LLMs' ability to perform complex code-related tasks effectively.

With \textbf{RQ2}, we explored the impact of inaccurate question on LLM comprehension of legacy code. While minor inaccuracies in comments have negligible effects on LLM comprehension, major inaccuracies significantly degrade understanding. This degradation occurs even when the LLMs have prior knowledge of the code, highlighting their tendency to accept all input as accurate and attempt to reconcile discrepancies, sometimes leading to erroneous conclusions. These insights underscore the importance of accurate and comprehensive documentation, particularly when leveraging LLMs for tasks involving legacy codebases.

In conclusion, our research emphasizes the critical role of documentation quality in maximizing the efficacy of LLMs in software development and maintenance tasks. 

Although it may be apparent to readers that inaccurate comments degrade LLM comprehension of code, this study shows the extent to which LLMs can self-evaluate inaccuracies in comprehension when humans introduce errors into codebases, even if the LLM has encountered the material as part of its training corpus. As LLMs continue to be integrated into workflows involving legacy systems, ensuring the accuracy and completeness of code comments will be essential for achieving reliable and high-quality outputs. While this work and methodology demonstrate promise, we expect future work to enhance the generalizability and applicability of our findings.

\section{Future Work}

In contexts like code modernization efforts, where code may not be well documented, it is important to understand the impact of both comment prevalence and inaccurate comments on LLM comprehension. Degraded LLM comprehension may result in reduced quality of LLM outputs for related tasks. In this work, we identified the improvement of LLM comprehension with increased comment prevalence and the degradation of LLM comprehension with inaccurate comments. However, we recognize that these are preliminary findings that warrant more robust experimentation to further explore the impact of comments on LLM comprehension. 

In future work, we will be addressing three primary objectives: experiment with a larger corpus, automate multiple-choice quiz generation, and mitigate known LLM biases. In this experiment, our dataset consisted of a single assembly language module. In future iterations of the experiment, we plan to validate this methodology on larger, more complex codebases to evaluate its ability to scale across real-world systems and across multiple codebases and legacy languages. In the absence of an established multiple-choice benchmark for legacy languages, we developed our own quizzes with the help of assembly SMEs. However, as with all human evaluation, quiz creation is a time-intensive process \cite{meisner2024evalquiz}. We need to expedite the time required for reliable quiz generation in order for MCQA to be a viable LLM evaluation method for niche legacy languages. To accelerate quiz generation for a wider range of modules and languages, we plan to explore automated quiz generation and leverage LLMs to produce quizzes to evaluate LLM code comprehension. Lastly, the sensitivity of LLMs to answer positioning in MCQA is well-documented \cite{li2024can,wang2024llms}. In future work, we will experiment with randomizing the location of the correct answer to reduce the effect of LLM bias. 

\section*{Acknowledgment}

This work was supported by the MITRE Independent Research \& Development program. \copyright The MITRE Corporation. All Rights Reserved. Approved for Public Release; Distribution Unlimited. Public Release Case Number 25-0821.

\bibliographystyle{IEEEtran}
\bibliography{references}

\end{document}